\documentclass{acm_proc_article-sp}
\usepackage{bbm}
\usepackage{graphicx}
\usepackage{subfigure}
\usepackage{url}
\newcommand{\junk}[1]{}
\usepackage{multicol}
\usepackage{etoolbox}
\makeatletter
\patchcmd{\maketitle}{\@copyrightspace}{}{}{}

\newcommand{\squishlist}{ \begin{list}{$\bullet$} { \setlength{\itemsep}{0pt} \setlength{\parsep}{3pt} \setlength{\topsep}{3pt} \setlength{\partopsep}{0pt} \setlength{\leftmargin}{1.5em} \setlength{\labelwidth}{1em} \setlength{\labelsep}{0.5em} } } \newcommand{\squishlisttwo}{ \begin{list}{$\bullet$} { \setlength{\itemsep}{0pt} \setlength{\parsep}{0pt} \setlength{\topsep}{0pt} \setlength{\partopsep}{0pt} \setlength{\leftmargin}{2em} \setlength{\labelwidth}{1.5em} \setlength{\labelsep}{0.5em} } } \newcommand{\squishend}{ \end{list} }

\begin{document}

\title{A Consensus-Focused Group Recommender System }
%\title{A Group Recommender System For Consensus Decision Making }

\numberofauthors{3} %  in this sample file, there are a *total*
 
\author{
 Stratis Ioannidis \\
      \affaddr{Technicolor,Palo Alto}\\
       \email{stratis.ioannidis@technicolor.com}
\alignauthor
 S. Muthukrishnan \\
        \affaddr{Rutgers New Brunswick} \\
       \email{ muthu@cs.rutgers.edu}
 \alignauthor
 Jinyun Yan\thanks{Authors by alphabetical order.}  \\
       \affaddr{Rutgers New Brunswick} \\
       \email{ jinyuny@cs.rutgers.edu}
}
% There's nothing stopping you putting the seventh, eighth, etc.
% author on the opening page (as the 'third row') but we ask,
% for aesthetic reasons that you place these 'additional authors'
% in the \additional authors block, viz.
 
\date{25 January 2013}
% Just remember to make sure that the TOTAL number of authors
% is the number that will appear on the first page PLUS the
% number that will appear in the \additionalauthors section.

\maketitle
\begin{abstract}
In many cases, recommendations are consumed by groups of users rather than individuals. In this paper, we present a system which recommends social events to groups. The system helps groups to organize a joint activity and collectively select which activity to perform among several possible options.  We also facilitate the consensus making, following the principle of group consensus decision making.
Our system allows users to asynchronously vote, add and comment on alternatives.  We observe social influence within groups through  post-recommendation feedback during the group decision making process.  We propose a {\em decision cascading model}  and  estimate such social influence, which can be used to improve the performance of group recommendation.  
We conduct experiments to measure the prediction performance of our model. The result shows that the model achieves better results than that of independent decision making model. The demo is accessible at  \url{http://tinyurl.com/grouprecsys}.
%\url{http://74.95.195.230:8889/}.

\junk{
In many cases, recommendations are consumed by groups of users rather than individuals. In this paper, we build a system which recommends social events to groups. The system helps groups to organize a joint activity and collectively select which activity to perform among several possible options.  Besides generating recommendations suited for the group,  we follow the principle of group consensus decision making to design interfaces that facilitate the consensus making. Our system allows users to asynchronously vote, add and comment on alternatives.  We observe social influence within groups through  post-recommendation feedback during the group decision making process.  A decision cascading model is proposed to estimate such social influence, which can be used to improve the performance of group recommendation.  %We assume each member in the group will vote for an item based on his/her individual preference and decisions by other members. 
We conduct experiments to measure the prediction performance of our model. The result shows that the model achieves better results than that of independent decision making model. The demo is accessible at  \url{http://74.95.195.230:8889/}.}
\end{abstract}
%A category including the fourth, optional field follows...
%\category{D.2.8}{Software Engineering}{Metrics}[complexity measures, performance measures]


 % NOT required for Proceedings

\section{Introduction}
Existing recommender systems, for example, Netflix and Amazon, mainly target individuals. However, people engage in \emph{social} events as a group both offline and online, e.g., movie watching, online gaming, dining out, book reading, and so on. The inherent need for social events and the large content space have inspired recent interests in group recommendation. To date, a variety of early stage systems have been developed for different applications, such as group web page recommendation~\cite{letsbrowse},  vacations or tours to groups of tourists~\cite{ardissono2003},  music tracks and playlists to large groups of listeners~\cite{crossen2002}, and movies and TV programs to friends and family~\cite{polylens,zhiwen2006}.  

Prior work on group recommendation mainly focuses on the modeling of group preference. The intuition is that if we can infer group preference, the problem is reduced to individual recommendation.  PolyLens~\cite{polylens}  uses a ``pseudo-user'' to represent the group's preference.  The ``pseudo-user'' is either  created manually by group members or automatically merging the rating history of the group members. Besides averaging individual ratings, the paper~\cite{groupmodeling} analyzed many other aggregation strategies, such as least misery, fairness, plurality voting and so on. Amer-Yahia \emph{et al.}~ \cite{Amer-Yahia2009} defined a consensus function of the group which combines the aggregation and variance of individual ratings. Another approach is first obtaining the results of individual recommendations, then merging them to a group list which minimizes its distortion to individual lists. Baltrunas \emph{et al.} ~\cite{linas2010} applied several rank aggregation methods to merge recommendation results of individuals. 

Unfortunately, the post-recommendation behavior is usually neglected. In reality, when a group of people participates a joint event,  individuals collectively make the choice over a set of alternatives of which activity to perform.  Members and the social group itself possess social influence which contributes to the outcome, which is often different from those made by individuals. The group decision making process includes the interaction among members. Members get to know what other members like and dislike, and may decide either to conform to other members' preference or persuade them.
As opposed to individual recommendation, the recommender system can easily figure out a person's reaction to the recommendation by a single click or a score. We argue that group recommender system should not ignore this process to learn the group and the members' feedback to group recommendation. Such feedback can be used to measure the recommendation and improve the accuracy of the profiling of the group.  

\begin{figure}[!ht]
 \label{fig:framework}
 \includegraphics[width=0.4\textwidth, height=0.27\textwidth]{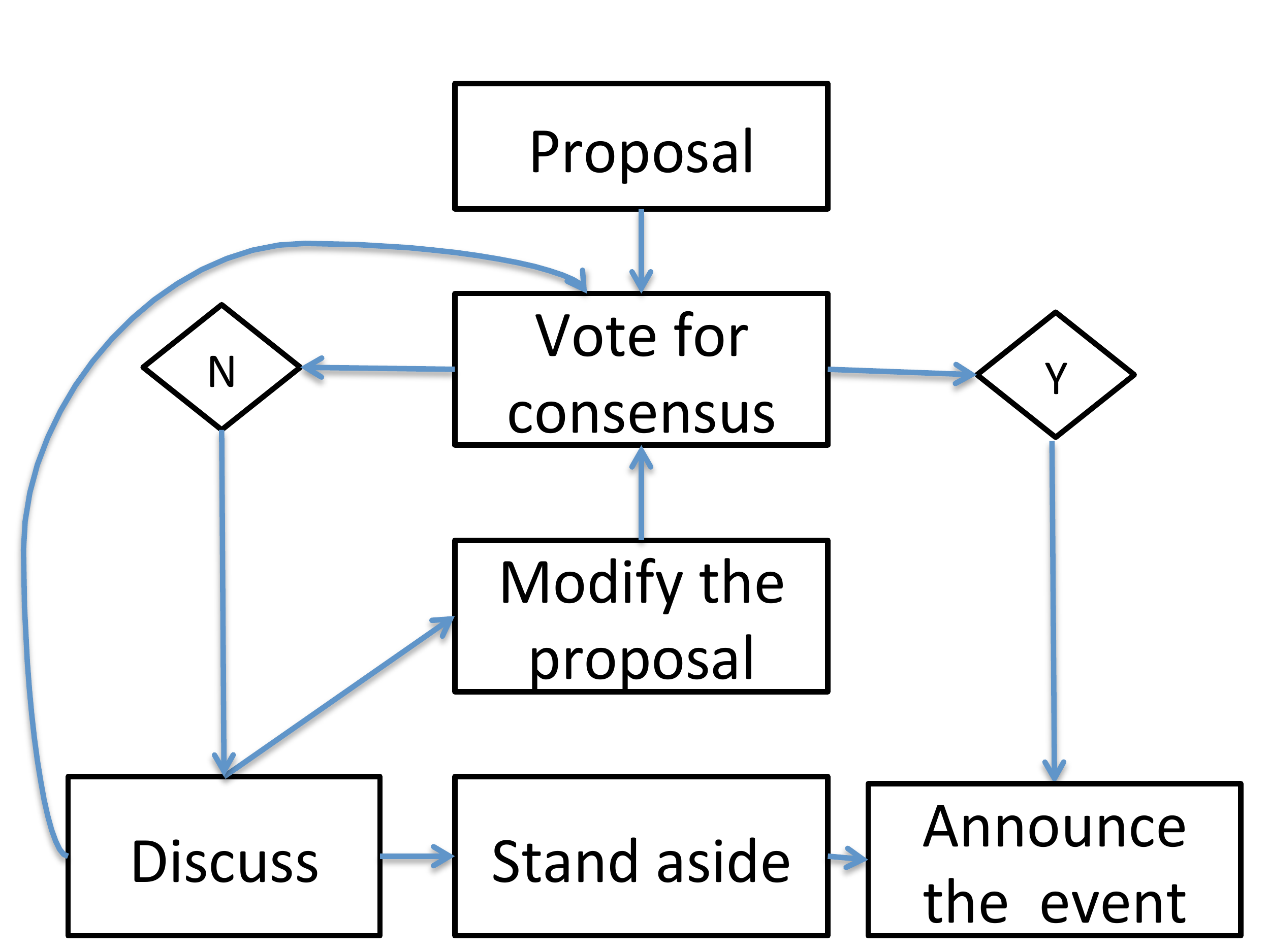}
 \caption{Group Decision Making }
  %\vspace*{-0.2cm}
 \end{figure}

Existing systems either only display group recommendation results to each individual as in ~\cite{polylens, groupfun}, or assume members can discuss alternatives offline~\cite{cats}. In our system, we display group recommendation results to each individual, and also allow members to discuss and decide on alternatives either simultaneously or asynchronously. We adopt voting as the consensus decision making mechanism. The interface design of the system follows the principle of group consensus decision making~\cite{consensus-decision-making}, and the framework of the group decision is in Figure~\ref{fig:framework}. The recommendation algorithm will contribute on proposal phrase, that occurs both before and during the group decision making. 
 
Members can vote in arbitrary order. When a member enter the decision process, other members may have already decided their preference. The previous decisions may influence the member. Social influence occurs when one's decision is affected by others. We observe social influence through the individual feedback during the discussion process.  An individual votes positively for an item according to his/her own preference on the item and other members' decisions. In this paper, we propose a decision cascading model to estimate the social influence among members.  We conducted experiments which show that using social influence can better predict the individual feedback in the voting process. We then propose the influence model to rank items for the group. 
 
The major contributions of our work are as follows.
%\vspace*{-0.5cm}
%\begin{itemize}
\squishlist
\item{ We build a group recommender system taking dining out events as the application. The system can generate recommendations suited for groups.}
\item{ We design an interface that supports asynchronous voting, adding and commenting on alternatives. }
\item{ We infer the social influence among members through their feedback in the consensus discussion process. }
\squishend
%\end{itemize} 

\section{The demo system}
\begin{figure*}[!ht]
  \begin{center}
   % \subfigure[Sign up]{\includegraphics[width=0.3\textwidth, height=0.2\textwidth]{photos/signup.pdf}}
    %\subfigure[Profile Page]{\includegraphics[width=0.35\textwidth,height=0.2\textwidth]{photos/mainpage.pdf}}
     \subfigure[Event Detail]{\includegraphics[width=0.47\textwidth,height=0.28\textwidth]{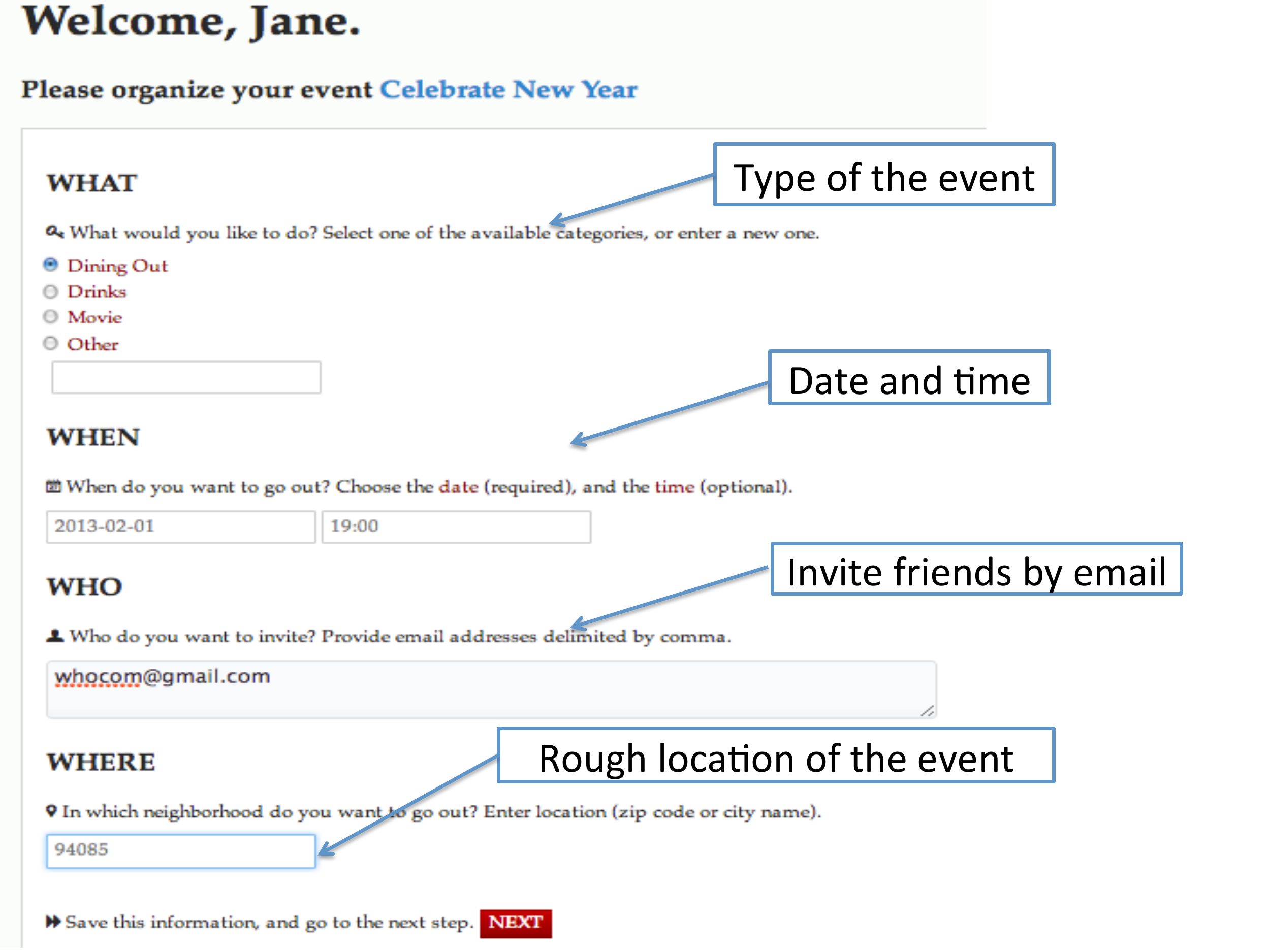}}
      \subfigure[Recommendation]{\includegraphics[width=0.47\textwidth,height=0.28\textwidth]{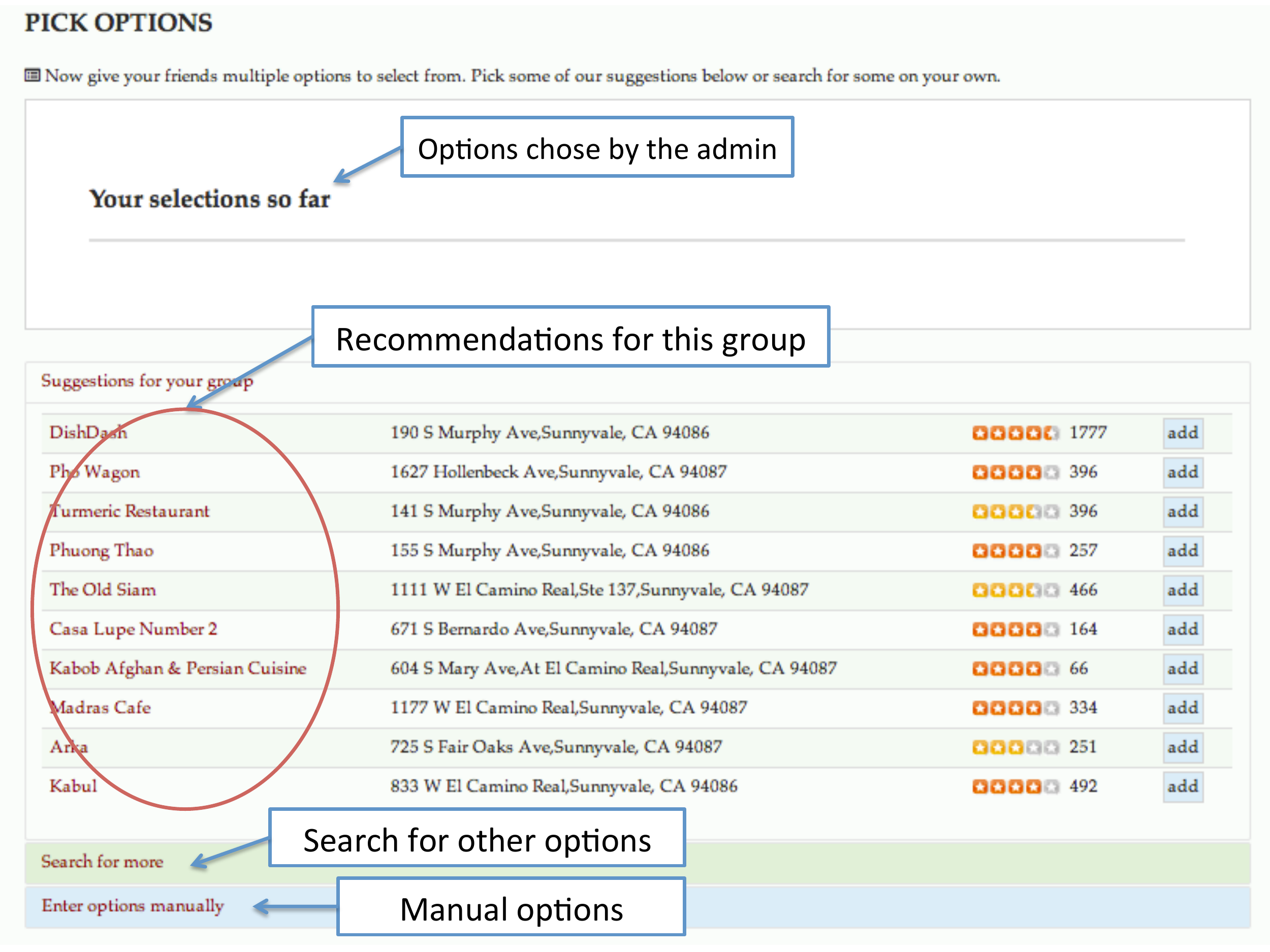}}
      \\
      \subfigure[Voting ]{\includegraphics[width=0.47\textwidth,height=0.28\textwidth]{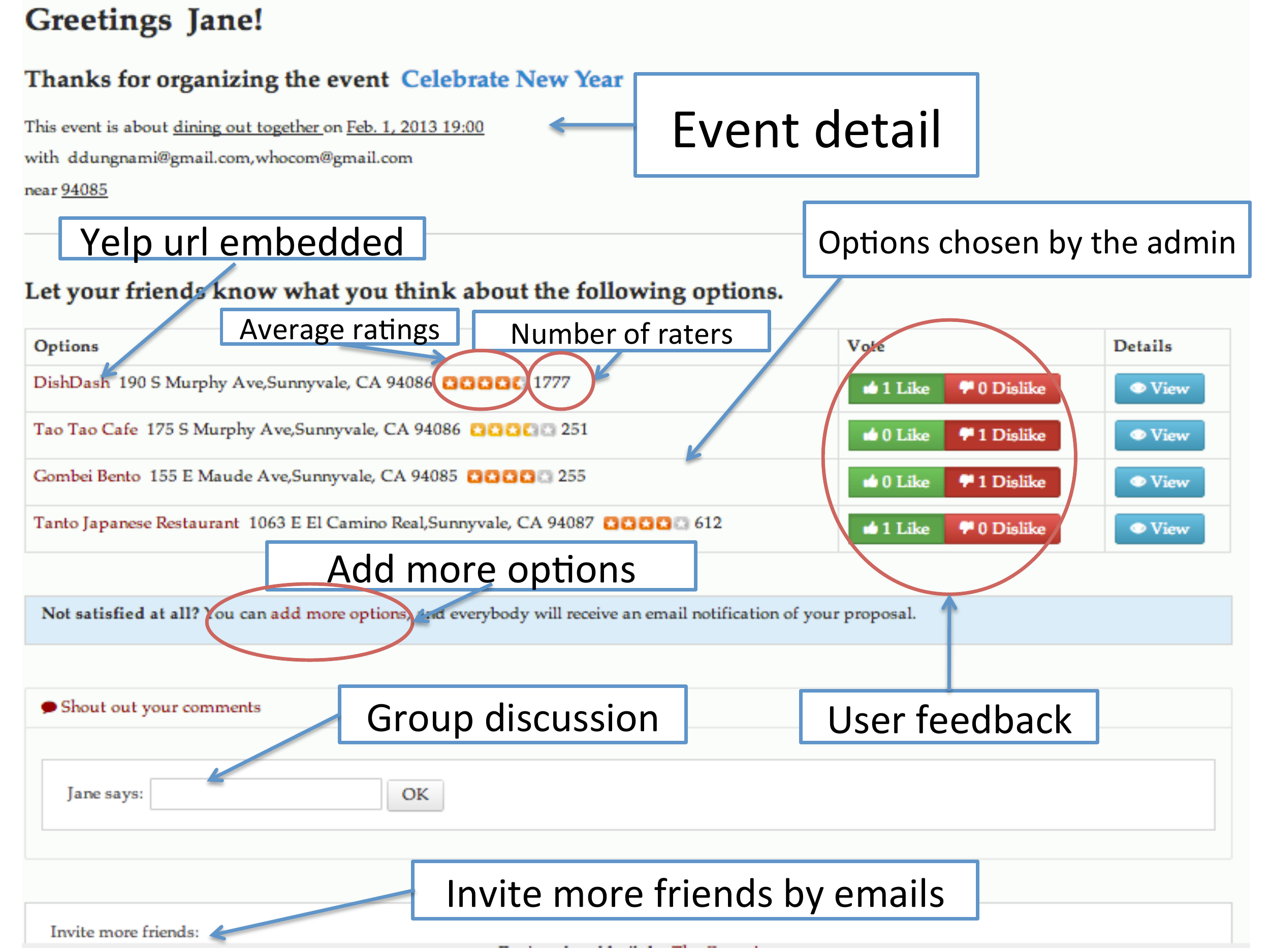}}
       \subfigure[View Detail ]{\includegraphics[width=0.47\textwidth,height=0.28\textwidth]{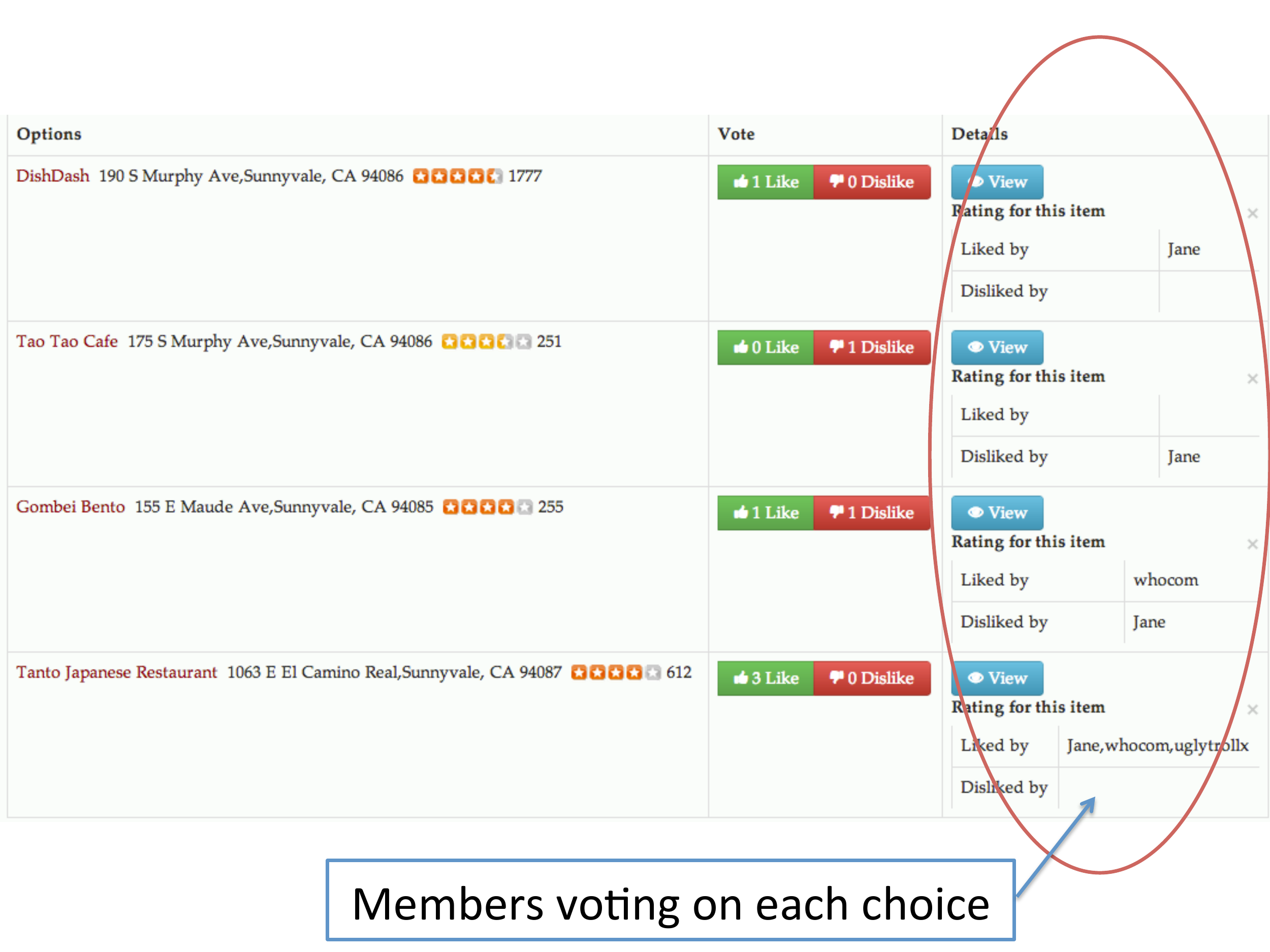}}
  \end{center}
%\vspace*{-0.2in}
  \caption{ Main Interfaces of the system}
  \label{fig:interfaces}
%\vspace*{-0.1in}
\end{figure*}

%We instantiate our group recommender system using dining out as the target application. 
We present our group recommender system. The demo is accessible at \url{http://74.95.195.230:8889/}. Figure~\ref{fig:interfaces} shows the screenshots of important interfaces.  A new user needs to sign up with a valid email address, because our system will send notifications through emails. When the user logs in the system, he/she will be presented the profile page, which includes creating a new event, displaying the current ongoing events, reviewing past events which are terminated and a friend list. The user who creates the event severs as the admin of the event. The admin will need to provide event details as in Figure~\ref{fig:interfaces}(a) including the category of the event, invited friends, the date and time of the event  as well as its location. The system recommends 5 friends to users, which is based on the frequency of co-participated events. These details except friends are used to filter candidates. Friends invited and the admin define the group participating the event.

 \junk{Figure~\ref{fig:interfaces}(a) is the sign up page for each user. Email is required such that the user can receive notifications. The user then can login the system using the email and his password.  Figure~\ref{fig:interfaces}(b) shows the profile page of the user, where allows the user to create a new event,  check the ongoing events which are not closed yet,  review the past events and see his friends list. The user who creates the event serves as the admin of the event.  The admin provides some detail of the events as in Figure~\ref{fig:interfaces}(c) including the category, the date and time, the location and email addresses of friends. The category,the date and time, and location are used for our recommender system to filter candidates. Friends plus the admin define the group.} 
Based on this context the system generates a recommended list with 10 items tailored for this group as in Figure~\ref{fig:interfaces}(b).  Restaurants are ranked by the prediction of group score. The admin can select one or more options from our recommendation list or alternatively search yelp from our interface. Manual entries are also possible.  After that, the admin will send out invitation to the group. Each participant in the group is called an attendee. The attendee will receive an email with a link to the voting page as in Figure~\ref{fig:interfaces}(c).

We here describe how we designed these interfaces to meet the principle of consensus decision making~\cite{consensus-decision-making}. 
%\vspace*{-0.2in}
\paragraph{Collaborative and Egalitarian}  All members of the group contribute to a shared proposal and have the opportunity to present, and amend proposals.  Thus, it is important that every member has the chance to propose restaurants to serve as options. The admin can select options during the event creation stage. All attendees including the admin can add more options on the voting page as in Figure~\ref{fig:interfaces}(c). The interface for attendees adding new options is similar to the one in Figure~\ref{fig:interfaces}(b). However, the recommended items will change if some users provided their feedback on some items already. 
%\vspace*{-0.2in}
\paragraph{Inclusive and Participatory} In a consensus process all group members are included and encouraged to participate. In our system, every member will receive an email notification of the event and be asked to make decisions on options. The voting page does not require registration before access. We adopt the URL encoding method to create a special link for each user and each event. Users are uniquely identified by their emails.  Non-registered users can also participate in the process. Whenever some members add more options, we will send an email to the group and ask them to provide feedback on the new options. Because members make decisions asynchronously, we provide a group chatting channel on the voting page. Members can leave messages and discuss on options.
%\vspace*{-0.2in}
\paragraph{Cooperative and Agreement Seeking} Participants in an effective consensus process should strive to reach the best possible decision for the group and all of its members, rather than competing for personal preferences. We display the number of positive and negative votes of each option in a real-time fashion.  We clearly and immediately display the voting results in order to emphasize the goal of achieving high agreement on an option.  The admin is the only person who can terminate the event and announce the final decision. When the admin terminated the event, every member will receive a summary of the event including the detail of options, votes and comments. The process and the result are  transparent to every member, which avoids unfair bias to certain individuals. 

We provide several types of information to each attendee to help them make decisions.  
 %\vspace{-0.2cm}
%\begin{enumerate}
\squishlist
\item Features of the item: below the title of each restaurant, we embed a Yelp link. Users can click it and check the details. We also provide the address, Yelp rating, and the number of raters of the place if exist. Being presented first and close to a restaurant, the average rating and the number of Yelp raters are likely to make  the strongest impression on users. 
\item Current votes: we display the accumulated positive votes and negative votes each option received so far. When a user provides his/her feedback on an option, the number of votes of this option changes correspondingly in real time.
\item Members' choice: users can view who voted what on each option as in Figure~\ref{fig:interfaces}(d). A person is affected by his friends' actions with different weight. Some members may have higher influence on a user than other members.  Thus we disclose the voting detail and help users decide whose decision to follow.  
\squishend
%\vspace{-0.2cm}
%\end{enumerate}
Recall that our goal of the voting page is to provide an fair and convenient environment to do consensus decision making. In reality, when the group wanted to make such decisions, they may call each other and discuss the options, send many rounds of emails,  or even organize a remote or face to face group meeting. All these efforts are to figure out what each member likes and dislikes and what options will be agreeable to all members. Through our system, we provide a place that members can asynchronously discuss options,  get to know other members' likes and dislikes, and conform to whoever is important to him/her. This information is  collected and displayed in one page, and helps avoiding the burden of group discussion. 

At last, the organizer needs to close the discussion and announce the final decision of the group by email notification. We rank  options by the number of positive votes and populate the top one automatically.  However, we still allow the organizer change the final decision to accommodate any possible change or offline discussion.
 
A system with new users will usually encounter the cold start problem. For this reason, our demo adopts content based algorithm.  Every user has a history of voted restaurants.  We extract features of each restaurant from Yelp, and predict the individual preference by its similarity with restaurants voted before. The group score is computed by the aggregation and the variance of individual predictions: $r(G,i) = w_1 \sum_{u \in G} r_{u,i} + w_2 (1- var(r_i) )$. One restaurant may be voted many times in different events. The score of each restaurant is the running average of all votes it received by this user. This may cause the change of the predicted preference each time of a new vote. 

\section{Social Influence Inference} 
In what follows, we discuss how we collect data from our system which can be used to improve the performance of recommendation. In particular, we propose a model that captures how a user is influenced by decisions made by other users. We demonstrate how we can learn parameters of the model from our data, and also show our model outperforms the baseline prediction using logistic regression.
 
Ye \emph{et al.}~\cite{ye2012} used a generative model to estimate the influence between two members.  It assumes that a person first chooses friends with certain probability, then friends choose items. However,  two users with similar interests consuming the same set of items will have high influence on each other.  We argue that during the negotiation process, we can observe the influence between members not relying on the similarity of the interests. A user may vote for an item not because she likes that item but her friends like it a lot. Such influence will cause conformity within the group.  We will describe the formal setting of the consensus negotiation process, then demonstrate how we leverage the feedback  during this process to infer the social influence among members.
 
% \paragraph{Decision Cascading Model}
\subsection{Decision Cascading Model}
Consider the scenario that a group $G$ with $|G|=d$ finite number of members tries to decide on the item $i$. During the consensus stage, every individual provides his/her own feedback to the item, which is represented by a binary value $y_{u,i} \in \{0,1\}$. Note that $y_{u,i}$ can be multinomial variable such as ``like'', ``dislike'' or ``neutral'', and it can also be a real number scaling from $1$ to $5$. We here only deal with the binary case.  Denote the group decision process by the group $G$ to the item $i$ by an event $e = (G, i, \{y_{u,i}\}_{u \in G})$. The decision of the item for the group depends on the number of positive votes. In other words, the average strategy is used to determine the group preference of an item as $y_{G,i}  =  1/|G| \sum_{u \in G} y_{u,i}$. 

We use $p(u | i)$ to denote the individual inherent preference on the item, $p(v | u)$ be the influence of the decision by the member $v$ to the member $u$. We now model how each individual votes given he/she is aware of existing decisions of others. The order with which people vote can be arbitrary and depends on when they access the decision process. Moreover, their vote can be influenced by the votes of other people voted before them.  Let $\mathcal{U}_{e}^+(u) $ be users who give positive feedback in event $e$ before user $u$ and $\mathcal{U}_{e}^-(u)$ be users who give negative feedback.  The probability that the member $u$ will give positive feedback to the item depends on his own preference on the item and other members' feedback as in~\eqref{influencemodel}.  Putting differently, a user $u$ votes positively by tossing $ |\mathcal{U}_{e}^+(u) | + 1$ independent $0-1$ coins and observing whether any of them returning $1$. The first coin is $1$ with probability $ p(u | i) $, and others with probability $p(v | u)$.
%\[  \Pr( y_{u,i} = 1 ) = 1 - (1- p(u | i ) ) * \prod_{v \in \mathcal{U}_e^+(u)} (1-p(v |u))   \]
 %\vspace*{-1cm}
\begin{equation}
\label{influencemodel}
 \Pr( y_{u,i} = 1 ) = 1 - (1- p(u | i ) ) * \prod_{v \in \mathcal{U}_e^+(u)} (1-p(v |u))  
\end{equation}
\vspace*{-.7cm}

 \subsection{Parameter Estimation}
%\paragraph{Parameter Estimation}
Given a set of events $E$, we can estimate the pairwise influence $p(v|u)$ and independent preference $p(u | i)$ by maximizing the likelihood of all events which the user $u$ has participated.
\begin{equation}
\label{Leu}
L(E,u) = \prod_{e:y_{u,i}=1} \Pr(y_{u,i}=1)\prod_{e:y_{u,i}=0} \Pr(y_{u,i}=0)
\end{equation}
We change the variables and let 
$1-p(u|i) = q_{u,i}$ ,
$ r_{u,i} = \Pr(y_{u,i}=1) $,
$ b_{v,u} = 1- p(v|u)$.
The problem becomes
\begin{align*}
\label{eq:gp}
\textrm{Maximize }   & \prod_{e:y_{u,i}=1} r_{u,i}  \prod_{e:y_{u,i}=0} \prod_{v \in U_e^-(u)} b_{v,u} q_{u,i} \\
 \textrm{ subject to }    \, & 0 \leq r_{u,i} \leq 1;  0 \leq q_{u,i} \leq 1    \, \forall i   \\
& 0 \leq b_{v,u} \leq 1    \, \forall v \\
%& 0 \leq q_{u,i} \leq 1  \textrm{   \forall i } \\
& r_{u,i} + \prod_{v \in U_e^+(u) } b_{v,i} q_{u,i} \leq 1    \, \forall e   \\
\end{align*}
 
Note the last constraint is an inequality rather than an equality. The objective function will strictly increase when either increase $r_{u,i}$ or $b_{v,i}$ or $q_{u,i}$, so the inequality will always be a binding constraint at the solution. The objective function is a monomial thus our problem is a geometric program which can be solved efficiently( see~\cite{myers2010}).  

The individual preference $p(u|i)$ reflects the interests of the individual and does not vary to a given event or a given group. We can estimate the individual preference from external resources, or  by selecting events that the individual is the first one to make decision and using logistic regression on these events. Then $q_{u,i} = 1-p(u|i)$ is treated as known constants when solving~\eqref{Leu}.   

 \subsection{Evaluation}
%\paragraph{Evaluation}
We use an offline evaluation to measure the performance of our decision cascading model. The basic idea is that if there is no pairwise influence on the decision, we can get good prediction on the individual feedback only using individual preference independently. Otherwise, our decision cascading model in~\eqref{influencemodel} will achieve better prediction performance. 
%The baseline algorithm ranks item by the average of individual prediction scores, and the individual score is obtained from logistic regression using movie features. 
We sampled 277 group events organized through our system. In total 19 individuals are involved, and 79 items are chosen. Groups have size 2,4 and 8.  Every user participated in 54 events on average.  For each event, members provide thumb up or thumb down feedback. A few of members did not provide any feedback. We collected 4398 valid votes at last. 
%We rank items by the number of positive votes it received as the ground truth ranking. 
We split events chronologically into training( 80\%)and testing set (20\%). The baseline algorithm uses logistic regression to predict individual preference.
\begin{table}[ht]
\centering
\begin{tabular}{|c|c|c|}
\hline
&  baseline &influence model \\
\hline
%below is average across users
true positive rate & 0.52 & 0.77 \\
false positive rate &0.35  & 0.6\\
% for first vote only, the train accuracy is 0.7586 the gap is quite big.
accuracy & 0.65  &0.7 \\
\hline
%this is average over all instances. 
AUC &  0.61  &0.85\\
\hline
\end{tabular}
\caption{Prediction performance on test set, prediction threshold=0.5 for accuracy}
\label{tb:compare_influence}
\end{table}
%\vspace*{-0.1in}

Table~\ref{tb:compare_influence} shows that the influence model has high true positive rate, better accuracy and much better AUC on the test set. However, the false positive rate is higher. This is caused that in our system, we actually provided three options, since some participants usually do not click on the ``Like'' nor ``Dislike''. When we separately deal with negative votes, we can get similar performance improvement. In the future work, we can change the interface to collect only binary decisions. On the other hand, we can work on the model with multinomial decision values.

\section{Conclusions}
We designed a group recommender system which not only generates and presents recommendation suited for the group, but also facilitates the consensus decision making of the group. We collected events data through our system and observed feedback manifested in the consensus decision making process. We inferred the social influence within the group by our decision cascading model. The experiment results show that our model has better prediction performance. In the future, we will work on the recommendation algorithms using the social influence, and compare with existing algorithms.
 
\bibliographystyle{abbrv}
\bibliography{grec}  % sigproc.bib is the name of the Bibliography in this case
% ACM needs 'a single self-contained file'!
%\balancecolumns
 
% That's all folks!
\end{document}